\pdfoutput=1
\documentclass[aps,prl,reprint,superscriptaddress,showpacs]{revtex4-1}
\usepackage{CJK}
\usepackage{graphicx}
\usepackage{dcolumn}
\usepackage{bm}
\usepackage{amssymb,amsmath}

\begin{document}

\title{Tunable terahertz radiation from graphene induced by moving electrons
}
\author{T. R. Zhan}
\email{phystrzhan@gmail.com}
\affiliation{Department of Physics, Key laboratory of Micro and Nano Photonic Structures
(Ministry of Education), and Key Laboratory of Surface Physics, Fudan
University, Shanghai 200433, P. R. China}
\affiliation{Bartol Research Institute, University of Delaware, Newark, Delaware 19716,
USA}
\author{D. Z. Han}
\affiliation{Department of Applied Physics, College of Physics, Chongqing University,
Chongqing 400044, P. R. China}
\affiliation{Department of Physics, Key laboratory of Micro and Nano Photonic Structures
(Ministry of Education), and Key Laboratory of Surface Physics, Fudan
University, Shanghai 200433, P. R. China}
\author{X. H. Hu}
\affiliation{Department of Materials Science and Laboratory of Advanced Materials, Fudan
University, Shanghai 200433, P. R. China}
\author{X. H. Liu}
\affiliation{Department of Physics, Key laboratory of Micro and Nano Photonic Structures
(Ministry of Education), and Key Laboratory of Surface Physics, Fudan
University, Shanghai 200433, P. R. China}
\author{S. T. Chui}
\affiliation{Bartol Research Institute, University of Delaware, Newark, Delaware 19716,
USA}
\author{J. Zi}
\email{jzi@fudan.edu.cn}
\affiliation{Department of Physics, Key laboratory of Micro and Nano Photonic Structures
(Ministry of Education), and Key Laboratory of Surface Physics, Fudan
University, Shanghai 200433, P. R. China}
\date{\today}

\begin{abstract}
Based on a structure consisting of a single graphene layer situated on a
periodic dielectric grating, we show theoretically that intense terahertz
(THz) radiations can be generated by an electron bunch moving atop the
graphene layer. The underlying physics lies in the fact that a moving
electron bunch with rather low electron energy ($\sim$1~keV) can efficiently
excite graphene plasmons (GPs) of THz frequencies with a strong confinement
of near-fields. GPs can be further scattered into free space by the grating
for those satisfying the phase matching condition. The radiation patterns
can be controlled by varying the velocity of the moving electrons.
Importantly, the radiation frequencies can be tuned by varying the Fermi
level of the graphene layer, offering tunable THz radiations that can cover
a wide frequency range. Our results could pave the way toward developing
tunable and miniature THz radiation sources based on graphene.
\end{abstract}

\maketitle

Graphene photonics and optoelectronics have attracted intense research
interest in recent years \cite{bon:10}. This is because graphene possesses
exceptional electronic and optical properties due to its unique electronic
band structure, i.e., the existence of Dirac cones \cite{gei:07}. Indeed, a
variety of novel applications such as broadband photodetectors, optical
modulators, and ultra-fast lasers have been proposed \cite{bon:10}.
Interestingly, graphene can support plasmons with frequencies in terahertz
(THz) and mid-infrared regimes \cite{wun:06}. Compared with surface plasmons
in noble metals \cite{bar:03}, GPs exhibit remarkable properties such as
deep subwavelength, extreme light confinement, and low Ohmic losses with a
further advantage of being tunable through electrostatic gating or chemical
doping \cite{jab:09,gri:12}. These features make graphene a promising
material for active plasmonic devices \cite{gri:12}, which could find
applications in transformation optics \cite{vak:11}, metamaterials \cite%
{ju:11}, and light harvesting \cite{tho:12} and concentrating \cite{dav:12}.

THz radiation with frequencies from 0.1~THz to 30~THz has attracted
increasing attention due to its wide range of potential applications \cite%
{ton:07}. However, a lack of desired sources of THz radiation limits the
realization of such applications. During the past decade, many approaches
including optically pumped solid-state devices, quantum cascade lasers,
diodes, and free-electron devices, have been investigated for the
development of THz sources \cite{gal:04}. Free-electron THz sources, wherein
radiation occurs as moving electrons interact typically with a perturbing
element \cite{smi:53,cok:09,osh:01}, are of particular interest owing to
their high power and continuous tunability by varying electron energies \cite%
{gal:04}. However, the difficulty in reducing their size while retaining
their broad tunability remains a great challenge for the applications of
such sources.

In this Letter, we show theoretically that intense THz radiations can be
generated by a moving electron bunch atop a graphene layer situated on a
periodic dielectric grating. We demonstrate the key role of GPs excited by
the moving electron bunch in the radiation process. In particular, we find
that the radiation intensity is strongly enhanced due to low losses and high
confinement of the excited GPs. The radiation patterns and frequencies can
be tuned by varying the electron velocity or the Fermi level of the graphene
layer. Our results may open up a new route to develop miniature and
tunable free-electron THz radiation sources based on graphene.

\begin{figure}
\vspace*{0pt} \centerline{\includegraphics[angle=0,width=8cm]{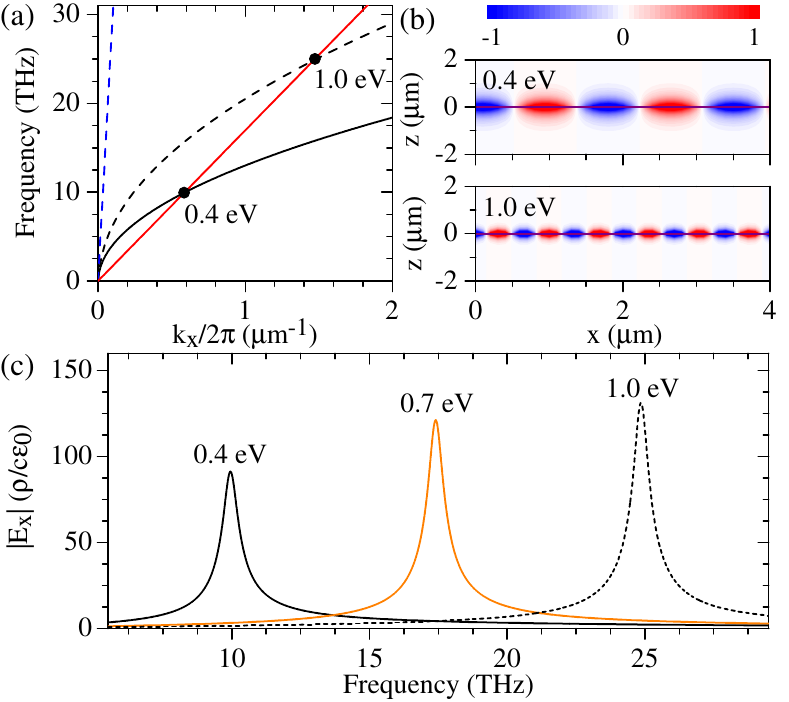}}
\vspace*{0pt}
\caption{(Color online) Excitation of GPs in a graphene layer
at $z=0$ on a dielectric substrate with $\varepsilon_d = 4$ by an electron
bunch moving with constant velocity $v$ in the $x$
direction and at $z=b$. The graphene has an intrinsic relaxation time of
$\protect\tau = 0.6$~ps throughout this work. (a) Dispersion curves of GPs
at different $E_{F}$ (black lines). The red solid line shows the electron
beam line $\protect\omega/k_x = v$ with $v = 0.057c$, and the blue dashed
line represents the light line $\protect\omega/k_x = c$, where $c$ is the
light speed in air. (b) Distribution of field Re($E_{x}$) for two GP
modes [dots in (a)]. (c) $\left\vert E_{x}\left(z=0\right) \right\vert$
versus frequency for $v =0.057c$ and $b=0.1$~$\protect\mu \text{m}$
at different $E_{F}$.}
\end{figure}

\begin{figure}
\vspace*{0pt} \centerline{\includegraphics[angle=0,width=8cm]{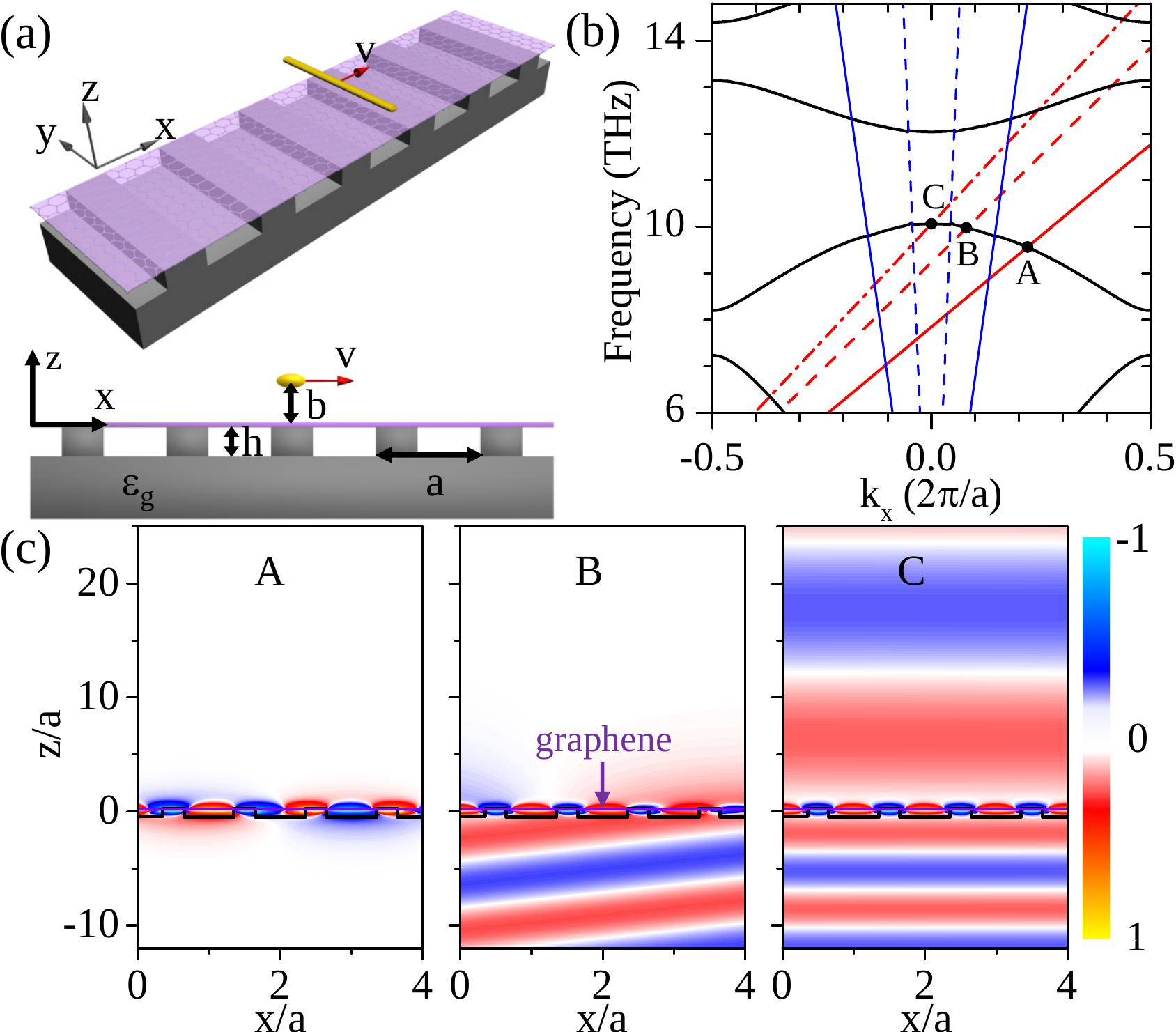}}
\vspace*{0pt}
\caption{(Color online) (a) 3D view (upper panel) and side view (lower panel) of an
electron bunch moving atop a graphene layer on an 1D dielectric grating. (b) Dispersion
curves (black solid lines) of GPs for the structure in (a). Red solid, dashed,
and dash-dotted lines show folded electron beam lines $\omega /(k_{x} + 2\pi/a)=v$ with
$v=$ $0.034c$, $0.04c$, and $0.0436c$, respectively. Blue dashed and solid lines represent
light lines $\omega /k_{x}=$ $c$ (air) and $\omega /k_{x}=c/\sqrt{\varepsilon_\text{g}}$ (substrate),
respectively.
(c) Distribution of field Re$\left(E_{x}\right)$ at frequencies of modes A, B, and C excited
by the electron bunch with corresponding $v$ in (b) and $b = 0.1$~$\mu \text{m}$. The black
solid lines in (c) depict the profile of the grating.}
\end{figure}

\begin{figure}[tbp]
\vspace*{0pt} \centerline{\includegraphics[angle=0,width=7.8cm]{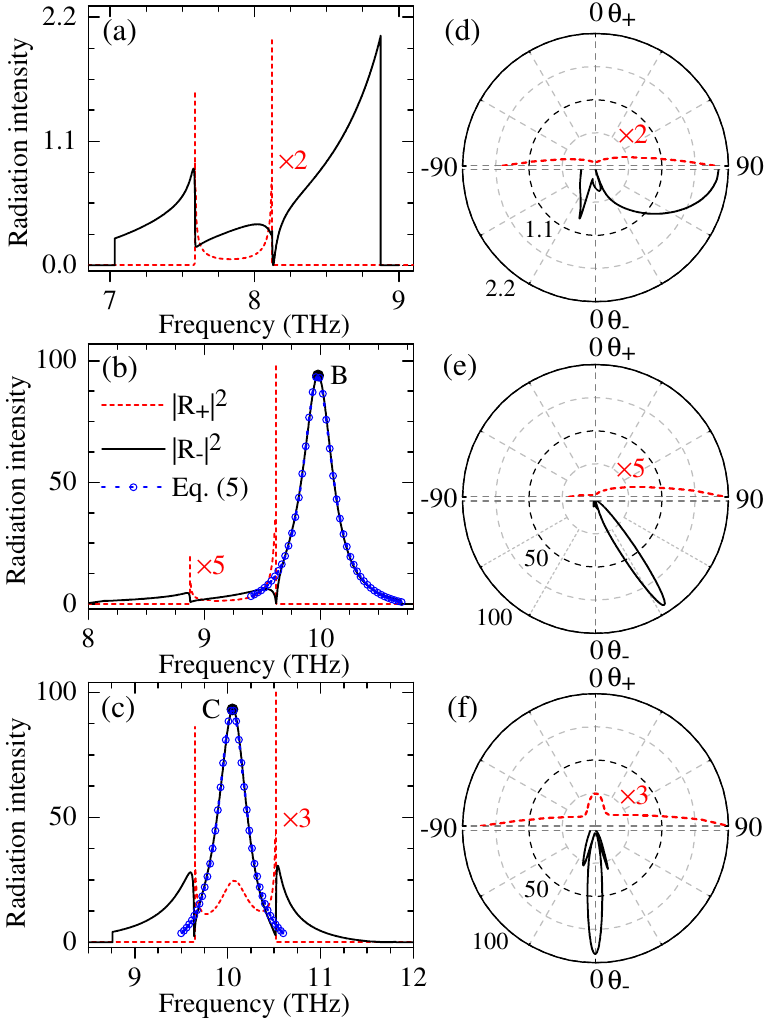}}
\vspace*{0pt}
\caption{(Color online) Radiation intensity $|R_{+}|^{2}$ toward the air side (red
dashed lines) and $|R_{-}|^{2}$ toward the substrate side (black solid lines) versus
frequency (left panels) and observation angle $\protect\theta$ (right panels)
for the setup shown in Fig. 2(a). Blue dotted lines with open circles represent the results of
fitting to Eq.~(5). The velocity of electron bunch is $v = 0.034c$
in (a) and (d), $0.04c$ in (b) and (e), and $0.0436c$ in (c) and (f). The modes
B and C in (b) and (c) are related to those in Fig.~2, respectively.}
\end{figure}

We start by considering an electron bunch moving atop a graphene layer on a
dielectric substrate. Suppose the electron bunch moves at a constant
velocity $v$ in the $x$ direction atop the graphene layer by a distance $b$.
For simplicity, the electron bunch is assumed to be uniform along $y$ with a
line charge density $\rho $. The moving electron bunch can be viewed as a
source with a current density $\mathbf{J}(\mathbf{r},t)=\hat{\mathbf{x}}\rho
v\delta (z-b)\delta (x-vt)$, which may induce a transverse-magnetic
electromagnetic (EM) wave of a form $e^{i(k_{x}x+k_{z}|z-b|)}$ with its
magnetic field polarized along $y$ in the frequency domain \cite{zha:s13},
representing a plane wave of a wave vector $k_{x}\hat{\mathbf{x}}+$sgn$%
(z-b)k_{z}\hat{\mathbf{z}}$, where $k_{x}=\omega /v$, $k_{z}=\omega \sqrt{%
1/v_{p}^{2}-1/v^{2}}$, and $\omega $ and $v_{p}$ are the angular frequency
and phase velocity of light in an ambient medium, respectively. Obviously,
Cherenkov radiation occurs when $v>v_{p}$ \cite{lan:84}, whereas no
radiation is expected when $v<v_{p}$ since $k_{z}$ is purely imaginary and
therefore the induced EM wave is evanescent.

Doped or gated graphene can support GPs that propagate along a graphene
layer with associated EM fields strongly confined near its surface \cite%
{wun:06,jab:09}, as shown in Fig.~1(b). GPs possess slow phase velocity \cite%
{jab:09}, which is approximated by $v_{\text{gp}}/c\simeq \frac{4\alpha }{%
1+\varepsilon _{d}}\frac{E_{F}}{\hbar \omega }$ in the nonretarded regime ($%
k_{\text{gp}}\gg \omega /c$) \cite{jab:09}, where $k_{\text{gp}}$ is the GP
wavevector, $E_{F}$ is the Fermi energy, $\varepsilon _{d}$  is the
permittivity of the dielectric substrate, $\alpha $ is the fine-structure
constant, and $c$ is the light speed in air. Note that $v_{\text{gp}}/c$ can
be of the order of $10^{-1}\sim 10^{-2}$ and further tuned by varying $E_{F}$
at THz frequencies, suggesting that graphene can be used as a tunable
slow-wave structure.

When the electron bunch moving atop a graphene layer, the induced evanescent
EM wave can excite GPs if satisfying the phase-matching condition \cite%
{gar:10}, namely
\begin{equation}
k_{\text{gp}}(\omega )=\omega /v.
\end{equation}%
Clearly, Eq.~(1) is equivalent to $v_{\text{gp}}(\omega )=v$, which is
satisfied at intersections of GP dispersion curves with electron beam lines.
Thus, the GP modes excited are given by $\omega _{\text{gp}}\simeq \frac{%
4\alpha }{1+\varepsilon _{d}}\frac{E_{F}}{\hbar v/c}$. As shown in
Fig.~1(a), for $E_{F}=0.4$~eV, the frequency of the excited GP is about $10$
THz with $v=0.057c$ (corresponding to rather low electron energy $0.832$%
~keV). The excited GP frequency can be tuned by adjusting $E_{F}$, e.g.,
about $24.9$~THz for $E_{F}=1.0$~eV.

To give a quantitative description, we solve rigorously Maxwell equations in
the frequency domain in order to obtain the EM fields induced by the moving
electron bunch. By considering the EM fields associated with the moving
electron bunch as incident fields upon the graphene layer \cite%
{van:73,yam:02}, reflection and transmission should be expected, yielding
the total electric fields as \cite{zha:s13}%
\begin{equation}
E_{x}(\mathbf{r},\omega )\equiv \zeta e^{ik_{x}x}\left\{
\begin{array}{ll}
e^{-\gamma _{0}|z-b|}-re^{-\gamma _{0}(z+b)}, & z>0 \\
t\eta e^{-\gamma _{0}b}e^{\gamma _{2}z}, & z<0%
\end{array}%
,\right.
\end{equation}%
where $\zeta =-i\frac{\rho }{2c\varepsilon _{0}}\frac{\gamma _{0}}{k_{0}}$, $%
r=\frac{1-\eta +\xi }{1+\eta +\xi }$ and $t=\frac{2}{1+\eta +\xi }$ are
reflection and transmission coefficients of the graphene layer respectively,
$\eta =\gamma _{2}/(\varepsilon _{d}\gamma _{0})$, $\xi =i\sigma _{\text{gp}%
}\gamma _{2}/(\varepsilon _{0}\varepsilon _{d}\omega )$, $\gamma _{0}=\sqrt{{%
k_{x}}^{2}-{k_{0}}^{2}}$, $\gamma _{2}=\sqrt{{k_{x}}^{2}-\varepsilon _{d}{%
k_{0}}^{2}}$, $k_{x}=\omega /v$, and $k_{0}=\omega /c$. At THz frequencies,
graphene conductivity simplifies to $\sigma _{\text{gp}}(\omega )=4\alpha
\epsilon _{0}c\frac{iE_{F}}{\hbar (\omega +i/\tau )}$ on the condition that $%
E_{F}\gg k_{B}T$, where $\tau $ is the relaxation time, $T$ is the
temperature, and $k_{B}$ is the Boltzmann constant \cite{jab:09}. Note that
poles of the reflection coefficient ($\eta +\xi =-1$) correspond exactly to
the phase matching condition of Eq.~(1), i.e., the excitation of GPs \cite%
{zha:13}. As a result, $E_{x}$ exhibits resonances due to the excitation of
GPs, as shown in Fig.~1(c), indicating that the EM fields near the graphene
layer are considerably enhanced around GP resonant frequencies. From
Eq.~(2), the enhancement factor is exactly $|t\eta |$, being about $\frac{2}{%
1+\varepsilon _{d}}\tau \omega _{\text{gp}}$ at GP resonant frequencies. For
example, at frequencies of $3$ and $30$~THz, $\tau \omega _{\text{gp}}$ can
be about $11$ and $110$ with $\tau \sim 0.6$~ps respectively \cite{zha:n3}.

As shown, an electron bunch moving atop a graphene layer can efficiently
excite GPs. However, the excited GPs cannot couple into free space due to
the wavevector mismatch between GPs and free-space radiations. To transform
GPs into free-space radiations, we consider a graphene layer situated on a
periodic dielectric grating, as schematically shown in Fig.~2(a). The
grating consists of 1D periodic grooves on a dielectric substrate. The
groove has period $a$, thickness $h$, and filling fraction $f$. In the
following discussions, the dielectric substrate has $\varepsilon _{\text{g}%
}=11.6$, and other parameters are $a=1.3 $~$\mu $m, $h=0.4$~$\mu $m, $f=0.7$%
, and $b=0.1$~$\mu $m.

The underlying physics for the transformation of excited GPs into free-space
radiations stems from the fact that the wavevector mismatch can be
compensated by reciprocal lattice vectors of the grating, namely
\begin{equation}
k_{\text{gp}}+2\pi m/a=\left\{
\begin{array}{ll}
k_{0}\sin \theta _{+}, & z>0 \\
\sqrt{\varepsilon _{\text{g}}}k_{0}\sin \theta _{-}, & z<0%
\end{array}%
,\right.
\end{equation}%
where $m$ is the diffraction order of the grating and $\theta $ is the
radiation angle with respect to $z$. Different branches of diffraction order
$m$ represent band foldings. Consequently, the GP dispersion is now
characterized by a well-defined band structure, as shown in Fig.~2(b), which
is obtained numerically by a scattering matrix method \cite{zha:12}.
Bandgaps appear at the Brillouin zone center and boundaries due to multiple
Bragg scatterings arising from the introduced periodicity. As a result, GP
modes can reside above the light line, and therefore radiate into free space
when excited by the moving electron bunch [Fig.~2(c)], giving rise to the
so-called diffraction radiation \cite{gar:10,ada:09,bak:09}. Note that there
are two kinds of light lines, one for the air side and the other for the
substrate side. The radiation patterns and directionality of excited GPs are
determined by their positions in the Brillouin zone. In this study, we focus
on GP modes in the first folded band ($m=\pm 1$); GP modes in other bands
can be analyzed similarly based on Eq.~(3).

\begin{figure}
\vspace*{0pt} \centerline{\includegraphics[angle=0,width=8cm]{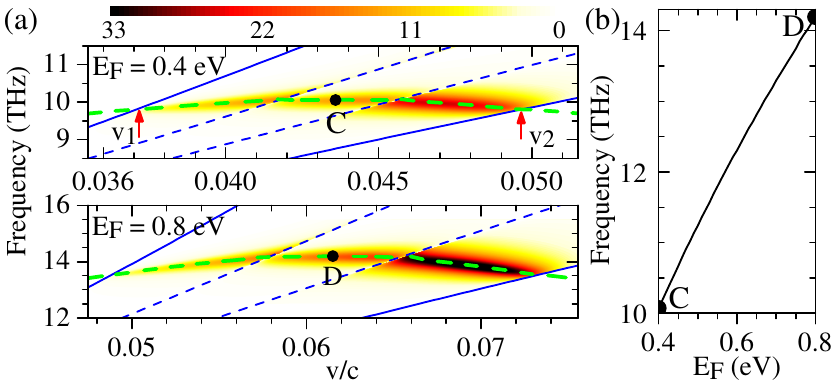}}
\vspace*{0pt}
\caption{(Color online) (a) Spectral density of radiated energy
$dW_{-}/d\omega$ toward the substrate versus $v$ and frequency for a setup
with $b = 0\text{ as }$shown in Fig. 2(a). $dW_{-}/d\omega $ is of unit
of $\rho^{2}/\epsilon _{0}c$ and plotted in color scale form. Green dashed
lines are obtained from Eq.~(1). Blue dashed (blue solid) lines are obtained
from intersection points of red lines with blue dashed (blue solid) lines in
Fig. 2(b). (b) Frequency of the excited GP at $k_x = 0$ [see Fig.~2(b)]
versus $E_F$. The corresponding $v$ can be calculated from $\omega/(k_x + 2\pi/a) = v$. }
\end{figure}

To characterize quantitatively the radiation induced by the moving electron
bunch, we obtain the spectral density of radiated energy as \cite{zha:s13}
\begin{equation}
\frac{d}{d\omega }{W}_{\pm }=\frac{\rho ^{2}}{4\pi \varepsilon _{0}c}\frac{%
\cos (\theta _{\pm })}{\sqrt{\varepsilon _{\pm }}}|R_{\pm }(\omega
)|^{2}e^{-2\gamma _{0}b},
\end{equation}%
where the plus (minus) sign stands for the air (substrate) side, and $R$ is
referred to as the radiation factor with $R_{+}$($R_{-}$) being the
reflection (transmission) coefficient of the structure \cite{van:73}.

Figures~3(a)-3(c) show the radiation intensity $|R_{\pm }(\omega )|^{2}$ as
a function of frequency. From Eq.~(3), the radiation intensity is also
plotted as a function of radiation angle $|R_{\pm }(\theta )|^{2}$, shown in
Figs.~3(d)-3(f). For $v=0.034c$ [Fig.~3(a)], since the excited GP mode [mode
A in Fig.~2(b)] lies below the light line of both air and substrate, it
cannot contribute to free-space radiations. Only a continuum of
Smith-Purcell (SP) radiation from the grating is observed toward both sides
\cite{smi:53,van:73}. Also, the radiation is distributed over a broad
angular range [Fig.~3(d)]. In contrast, for $v=0.0436c$, prominent resonant
peaks in the radiation spectra can be observed toward both the air and
substrate sides [Fig.~3(c)], corresponding exactly to the excited GP mode
[mode C in Fig.~2(b)]. Interestingly, the radiation is highly directional
and normal to the grating [Fig.~3(f)]. Note that there exists still SP
radiation toward both sides but with much lower intensity compared with that
from the excited GP. When $v=0.04c$ [Fig.~3(b)], a resonant peak positioned
at the excited GP frequency [mode B in Fig.~2(b)] is observed only toward
the substrate with its intensity dominating over that of SP radiation. From
Fig.~3(e), highly directional radiation at an oblique angle is observed
toward the substrate, whereas a broad angular emission of SP radiation
occurs toward air.

To gain a deeper insight into the physics of the strongly enhanced radiation
from excited GPs, we develop a self-consistent electromagnetic theory of the
coupling between moving electrons and GPs by assuming that the induced EM
fields can be expressed in terms of GP modes around resonant frequencies $%
\omega _{\text{gp}}$ \cite{zha:s13}. The theory can provide a closed-form
expression for the radiation intensity as
\begin{equation}
|R_{\pm }(\omega )|^{2}=F\frac{\omega _{\text{gp}}^{2}}{4Q^{2}(\omega
-\omega _{\text{gp}})^{2}+\omega _{\text{gp}}^{2}}\frac{Q}{Q_{r,\pm }}\frac{%
\sqrt{\varepsilon _{\pm }}}{\cos (\theta _{\pm })},
\end{equation}%
where $F=\frac{2}{\pi }\frac{Q}{V/\lambda }$, $V$ is the generalized mode
volume \cite{zha:n5}, $Q$ is the total quality factor, $Q_{r,\pm }$ is the
radiative quality factor associated with the coupling of GPs into free space
toward the air (substrate) side, and $\lambda =2\pi c/\omega _{_{\text{gp}}}$%
. Note that $|R_{\pm }(\omega )|^{2}$ exhibits a Lorentzian line shape,
reaching a maximum of $F\frac{Q}{Q_{r,\pm }}\frac{\sqrt{\varepsilon _{\pm }}%
}{\cos (\theta _{\pm })}$ at resonant frequencies. It is found that both $Q$
and $\lambda /V$ can be of the order of $10$ \cite{zha:n4}, and therefore
the radiation enhancement arises from high $Q$-factors and small mode
volumes of excited GPs.

To describe quantitatively the radiated energy, its spectral density as a
function of $v$ and frequency at different $E_{F}$ is shown in Fig.~4(a). We
only discuss the radiation toward the substrate since its intensity is much
larger than that toward air. For a given $E_{F}$, the radiation from excited
GPs occurs over a narrow frequency range, showing a weak dependence on $v$.
This weak dependence stems from the weak dispersion of GP bands above the
light line \cite{zha:13}. However, the radiation peak shows a strong
dependence on $E_{F}$. From Fig.~4(b), the peak frequency varies roughly
from $10$ to $14.2$ THz as $E_{F}$ increases from $0.4$ to $0.8$~eV. Note
that the spectral range of the radiation can be further tuned by engineering
the grating structure \cite{zha:s13}.

\begin{figure}
\vspace*{0pt} \centerline{\includegraphics[angle=0,width=8cm]{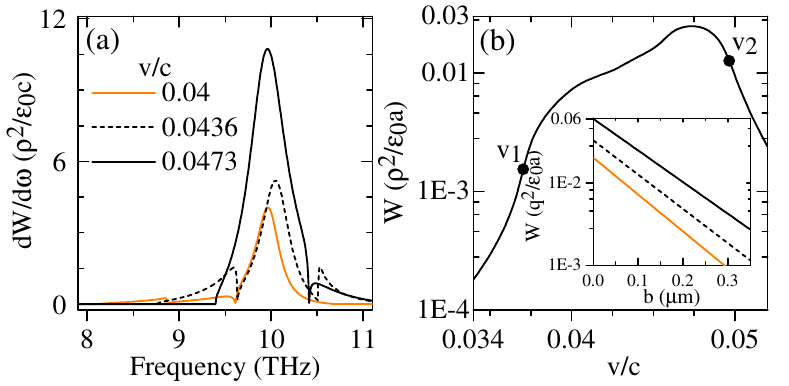}}
\vspace*{0pt}
\caption{(Color online) (a) Spectral density of radiated energy $dW_{-}/d\omega $
toward the substrate versus frequency, and (b) radiated energy $W_{-}$ toward the
substrate versus $v$ for a setup with $b=0.1$~$\mu \text{m}$ as shown in
Fig.~2(a). The inset to (b) plots the radiated energy $W_{-}$ versus $b$ for
different $v$ studied in (a). $v_1$ and $v_2$ in (b) correspond to those
in Fig.~4(a), respectively.}
\end{figure}

By integrating the spectral density of Eq.~(4) over frequency and radiation
angle, the total radiated energy $W$ can be obtained, shown in Fig.~5(b) as
a function of $v$. The radiated energy is dominate over the velocity range
where the excited GPs contribute to the radiation. For example, for $%
v=0.0436c$, the radiated energy is $W=1.26\times 10^{-2}$~$\rho
^{2}/\epsilon _{0}a$, nearly two order of magnitude larger than $%
W=1.81\times 10^{-4}$~$\rho ^{2}/\epsilon _{0}a$ for $v=0.034c$, where only
SP radiation occurs [Fig.~4(a)]. To evaluate quantitatively the radiated
energy, we consider an electron bunch with $\rho =100$~pC/cm. The total
radiated energy is estimated to be $W=1.099\times 10^{-3}$~$\mu \text{J/cm}%
^{2}$ for $v=0.0436c$, corresponding to a peak power of $1.31$~kW/$\text{cm}%
^{2}$ for a radiation pulse of $0.841$~ps \cite{zha:n2}. The peak value of $%
dW/d\omega $ is found to be $0.0123$~$\mu \text{J/cm}^{2}$/THz at $10.05$~$%
\text{THz.}$

In our calculations, the induced GP radiation is mainly influenced by the
impurity- and phonon-limited relaxation time $\tau =0.6$~ps \cite{zha:s13},
which is estimated for $E_{F}=0.4$~eV from the measured DC mobility $\mu
=1.5\times 10^{4}$~cm$^{2}$/Vs at room temperature \cite{hao:13}. While
impurity scattering is the dominant factor limiting $\tau $ in low-quality
graphene \cite{che:08}, $\tau $ can be improved in high-quality graphene
which has been reported to achieve high mobility values, an order of
magnitude larger than what is assumed in this work \cite%
{gei:07,bol:08,dea:10}. On the other hand, graphene optical phonons
significantly degrade $\tau $ for frequencies above $\omega _{\text{oph}%
}=48.4$~THz \cite{jab:09,yan:13}. Howerver, the THz frequency regime of
interest is below $\omega _{\text{oph}}$. Therefore, high-performance GP
radiation could be realized in the THz regime.

We now consider the experimental implementation of our proposal. The
fabrication of high-quality graphene and its integration with subwavelength
dielectric gratings have already been demonstrated experimentally \cite%
{hao:13,gao:13}. While sub-ps electron bunches required in the THz regime
can be obtained in the keV energy range with the state-of-the-art
development of ultrafast pulsed electron sources by employing femtosecond
lasers \cite{mcc:13}, continuous electron beams from low-voltage electron
microscopes could also be employed to realize the effect shown in this
letter \cite{ada:09}. By taking advantage of high resolution of electron
microscopes, electron beams can be directed parallel to the surface of a
system over a distance on the nanometer scale \cite{ada:09,kin:05}. Although
charge density in sub-ps electron bunches with good beam quality may be
limited, electron bunch trains with high repetition rate could be used to
further increase the radiation intensity \cite{kor:05}.

In conclusion, we have shown that GPs excited by uniformly moving electrons
with low electron energies can give rise to THz radiations with strongly
enhanced intensity due to low losses and high confinement of GPs.
Importantly, the radiation frequency can be tuned by varying $E_{F}$ via
electrostatic gating or chemical doping. In addition, the possibility of
using low-energy electrons could overcome the size limit of conventional
free-electron THz sources requiring high-energy electrons \cite{gal:04}.
Therefore, our results could open up the possibility of developing miniature
free-electron THz radiation sources with high tunability based on graphene
plasmonics.

This work was supported by the 973 Program (Grant Nos. 2013CB632701 and
2011CB922004). The research of J.Z. is further supported by the NSFC.


\begin{thebibliography}{99}
\bibitem{bon:10} F. Bonaccorso \textit{et al.}, Nat.Photon. \textbf{4}, 611
(2010).

\bibitem{gei:07} A. K. Geim \textit{et al.}, Nat. Mater. \textbf{6}, 183
(2007).

\bibitem{wun:06} B. Wunsch \textit{et al.}, New J. Phys. \textbf{8}, 318
(2006); E. H. Hwang and S. Das Sarma, Phys. Rev. B \textbf{75}, 205418
(2007).

\bibitem{bar:03} W. L. Barnes \textit{et al.}, Nature (London) \textbf{424},
824 (2003).

\bibitem{jab:09} M. Jablan \textit{et al.}, Phys. Rev. B \textbf{80}, 245435
(2009); F. H. L. Koppens \textit{et al.}, Nano Lett. \textbf{11}, 3370
(2011).

\bibitem{gri:12} A. N. Grigorenko \textit{et al.}, Nat. Photon. \textbf{6},
749 (2012).

\bibitem{vak:11} A. Vakil and N. Engheta, Science \textbf{332}, 1291 (2011).

\bibitem{ju:11} L. Ju \textit{et al.}, Nat. Nanotechnol. \textbf{6}, 630
(2011).

\bibitem{tho:12} S. Thongrattanasiri \textit{et al.}, Phys. Rev. Lett.
\textbf{108}, 047401 (2012).

\bibitem{dav:12} S. Thongrattanasiri \textit{et al.}, Phys.
Rev. Lett. \textbf{110}, 187401 (2013).

\bibitem{ton:07} M. Tonouchi, Nat. Photon. \textbf{1}, 97 (2007).

\bibitem{gal:04} G. P. Gallerano, and J. A. Valdmanis, in Proceedings of the
2004 FEL Conference (Trieste, Italy, 2004), pp. 216.

\bibitem{smi:53} S. J. Smith \textit{et al.}, Phys. Rev. \textbf{92}, 1069
(1953).

\bibitem{cok:09} A. M. Cook \textit{et al.}, Phys. Rev. Lett. \textbf{103},
095003 (2009).

\bibitem{osh:01} P. G. O'Shea \textit{et al.}, Science \textbf{292}, 1853
(2001).

\bibitem{zha:s13} See Supplemental Material at for further details on the
derivation of Eqs.~(2), (4), and (5), tunability of induced GP radiation by
engineering the grating, a discussion of the influence of the relaxation
time, and electrostatic effects on the moving electrons for electrostatic
doping.

\bibitem{lan:84} L. D. Landau, E. M. Liftshitz, and L. P. Pitaevskii,
\textit{Electrodynamics of Continuous Media} (Pergamon, 1984).

\bibitem{gar:10} F. J. Garc\'{\i}a de Abajo, Rev. Mod. Phys. \textbf{82},
209 (2010).

\bibitem{van:73} P. M. van den Berg, J. Opt. Soc. Am. \textbf{63}, 689
(1973).

\bibitem{yam:02} S. Yamaguti \textit{et al.}, Phys. Rev. B \textbf{66},
195202 (2002); F. J. Garc\'{\i}a de Abajo \textit{et al.}, Phys. Rev. B
\textbf{67}, 125108 (2003).

\bibitem{zha:13} T. R. Zhan \textit{et al.}, J. Phys. Condens. Matter
\textbf{25}, 215301 (2013).

\bibitem{zha:n3} At $\omega =\omega _{\text{gp}}$, $|E_{x}(z=0)\mathbf{%
|\simeq }\frac{\rho }{c\varepsilon _{0}(1+\varepsilon _{d})}\frac{\tau
\omega _{\text{gp}}}{k_{0}/\gamma _{0}}e^{-\gamma _{0}b}$. Note that $%
1/\gamma _{0}$ is the GP decay length along $z$, and $k_{0}/\gamma
_{0}\simeq v_{\text{gp}}/c$ is of the order of $0.1\sim 0.01$. Hence, the
excitation of GPs by moving electrons also benefits from the strong
confinement of GPs.

\bibitem{zha:12} T. R. Zhan \textit{et al.}, Phys. Rev. B \textbf{86},
165416 (2012).

\bibitem{ada:09} G. Adamo, \textit{et al.}, Phys. Rev. Lett. \textbf{103},
113901 (2009); G. Adamo \textit{et al.}, J. Opt. \textbf{12}, 024012 (2010).

\bibitem{bak:09} M. I. Bakunov \textit{et al.}, Opt. Express \textbf{17},
9323 (2009); S. Liu \textit{et al.}, Phys. Rev. Lett. \textbf{109}, 153902
(2012).

\bibitem{zha:n5} $V\equiv \frac{\int [\varepsilon _{0}\varepsilon (\mathbf{r}%
)|\mathbf{\tilde{E}}|^{2}+\mu _{0}|\mathbf{\tilde{H}}|^{2}]dxdz}{%
2\varepsilon _{0}|\int \mathbf{\tilde{E}}_{x}^{\ast }(z=0)e^{ik_{x}x}dx|^{2}}
$ where $k_{x}=\omega /v$ and $\mathbf{\{\tilde{E},\tilde{H}\}}$ is the
field distribution of GP modes \cite{zha:s13}. Note that both
transverse-magnetic polariztaion and strong confinement of GPs are necessary
for achieving a small $V$.

\bibitem{zha:n4} By fitting the numerical results to Eq.~(5) \cite{zha:s13},
it is obtained that, when $\tau =0.6$ ps, $Q=25.4$ $(20.5)$, $\lambda
/V=6.9\ (13.1)$, and $Q/Q_{r,-}=0.21\ (0.31)$ for $v/c=0.0436$ $(0.0473)$.

\bibitem{zha:n2} Pulse duration of the radiation is estimated from the
spectral bandwidth of radiation peaks.

\bibitem{hao:13} Y. Hao \textit{et al.}, Science \textbf{342}, 720 (2013).

\bibitem{che:08} J. H. Chen \textit{et al.}, Nat. Nanotechnol. \textbf{3},
206 (2008).

\bibitem{bol:08} K. I. Bolotin \textit{et al.}, Solid State Commun. \textbf{%
146}, 351 (2008).

\bibitem{dea:10} C. R. Dean \textit{et al.}, Nat. Nanotechnol. \textbf{5},
722 (2010).

\bibitem{yan:13} H. Yan \textit{et al.}, Nat. Photon. \textbf{7}, 394 (2013).

\bibitem{gao:13} W. Gao \textit{et al.}, Nano Lett. \textbf{13}, 3698 (2013).

\bibitem{mcc:13} A. J. McCulloch \textit{et al.}, Nat. Commun. \textbf{4},
1692 (2013); P. Hommelhoff \textit{et al.}, Phys. Rev. Lett. \textbf{96},
077401 (2006); B. J. Siwick \textit{et al.}, Science \textbf{302}, 1382
(2003).

\bibitem{kin:05} W. E. King \textit{et al.}, J. Appl. Phys. \textbf{97}
(2005).

\bibitem{kor:05} S. E. Korbly \textit{et al.}, Phys. Rev. Lett. \textbf{94},
054803 (2005).
\end{thebibliography}
\end{document}